\documentclass[aps, prx, reprint, amsmath, amssymb, superscriptaddress, floatfix]{revtex4-2}

\usepackage{natbib}
\usepackage{graphicx}
\usepackage{caption}
\usepackage{dcolumn}
\usepackage{bm}
\usepackage[svgnames]{xcolor}

\usepackage{tcolorbox}
\usepackage{enumitem}

\setlist[itemize]{itemsep=2pt, topsep=0pt, parsep=0pt, partopsep=0pt}
\usepackage{hyperref}
\hypersetup{
    colorlinks=true,
    linkcolor=blue,
    citecolor=blue,
    urlcolor=blue
}

\begin{document}
    \title{NEMO: A Framework for Nematic and Morphological Analysis of \texorpdfstring{\\}{ } Curved and Multi-layered Biological Surfaces}

    \author{Konstantinos Andreadis}
    \affiliation{Department of Biochemistry, University of Geneva, Geneva, Switzerland}
    \affiliation{Department of Genetics and Evolution, University of Geneva, Geneva, Switzerland}

    \author{Oriol Mañé Benach}
    \affiliation{Department of Biochemistry, University of Geneva, Geneva, Switzerland}

    \author{Claire A. Dessalles}
    \affiliation{Institut Lumière Matière, CNRS and Université Claude Bernard Lyon 1, Villeurbanne, France}
    \affiliation{Department of Biochemistry, University of Geneva, Geneva, Switzerland}

    \author{Lodovico Mazzei}
    \affiliation{Department of Genetics and Evolution, University of Geneva, Geneva, Switzerland}
    
    \author{Aurélien Roux}
    \affiliation{Department of Biochemistry, University of Geneva, Geneva, Switzerland}

    \author{Guillaume Salbreux}
    \email{correspondence to: \\ guillaume.salbreux@unige.ch; aurelien.roux@unige.ch}
    \affiliation{Department of Genetics and Evolution, University of Geneva, Geneva, Switzerland}

    \date{\today}
    
    \begin{abstract}
        Across biological scales, from cytoskeletal networks to whole tissues, orientational order and topological defects arise within complex three-dimensional geometries.
        However, quantifying nematic order---orientational order with head-to-tail symmetry---remains challenging: 2D projections introduce geometric distortions, while current 3D methods often struggle to resolve distinct nematic fields within curved or multilayer structures.
        Here, we introduce NEMO, a modular Python framework for the depth-resolved quantification of tangential nematic order and surface morphology.
        By reconstructing biological surfaces as triangulated meshes, NEMO projects curved intensity layers, extracts local nematic directors, and computes locally averaged nematic tensors within the tangent plane.
        The pipeline identifies topological defects and computes their topological charge by accounting for the Gaussian curvature of the underlying surface.
        Furthermore, NEMO quantifies tissue morphology through inter-surface distance and surface-fitted estimates of Gaussian and mean curvatures.
        We show the capacities of this framework using a synthetic nematic film on a vesicle and experimental actin organisation in \textit{Hydra}.
        By combining customisable projections with surface-constrained analysis, NEMO provides a unified framework for quantifying the interplay between orientational order and geometry across scales.
    \end{abstract}


    \maketitle
    \twocolumngrid
    
    \section{Introduction}\label{sec:introduction}
    From cellular vortices \cite{guillamat2022integer} to whole-body actin alignment in \textit{Hydra} \cite{maroudas2021topological,ravichandran2025topology}, nematic order---orientational alignment with head-to-tail symmetry---plays a central role in tissue morphogenesis. Topological defects, which are singularities in the orientation field, have been shown to act as organising centres for collective dynamics and morphogenesis of epithelial tissues \cite{saw2017topological, doostmohammadi2022physics, guillamat2022integer}. However, accurately measuring nematic order in highly curved or multilayer biological structures presents major challenges.
    
    Standard image analysis approaches typically rely on two-dimensional maximum-intensity projections or planar cross-sections \cite{keber2014topology}.
    While convenient, these methods inherently distort the orientation field, displace defect positions, and merge signals of closely spaced layers.
    Recent computational advances have begun to address these limitations.
    For instance, Eckert et al. developed a framework for nematic analysis on tissue surfaces of arbitrary geometry \cite{eckert2025}, Tan et al. analysed the dynamics and elongation of cells on the surface of rotating pancreas spheres using spherical projections \cite{tan2024emergent}, and Claussen et al. developed a \textit{Blender}-based tissue cartography tool \cite{claussen2025blender}.
    
    Despite these progresses, existing tools to resolve thin layers on curved surfaces, computing tangential nematic order, and in determining defect charge on curved surfaces, remain limited.
    Furthermore, there remains a need to couple the extracted nematic field with the surface morphology.
    
    To address these needs, we introduce NEMO, a modular Python framework for the mesh-based analysis of tangential nematic order, topological defects, and morphological features on curved surfaces. NEMO reconstructs surfaces from volumetric images as triangulated meshes and ``\textit{peels}" intensity layers - i.e. projects on a surface intensity measured at a fixed distance normal to the surface. It then extracts local tangential nematic directors, computes the local nematic tensor average, detects topological defects, and computes their topological charge. It also quantifies Gaussian and mean curvatures and the distance separating several surfaces, typically useful to extract epithelial thickness. NEMO supports large multichannel time-lapse TIFF hyperstacks, integrates with external software, and provides interactive 3D visualisation through Napari. Here, we describe the NEMO framework and demonstrate its capabilities on synthetic and experimental datasets.
    

    \section{Results}\label{sec:results}

    \subsection{The NEMO framework}\label{subsec:the-nemo-framework}
\begin{figure*}
        \centering
        \includegraphics[width=\textwidth]{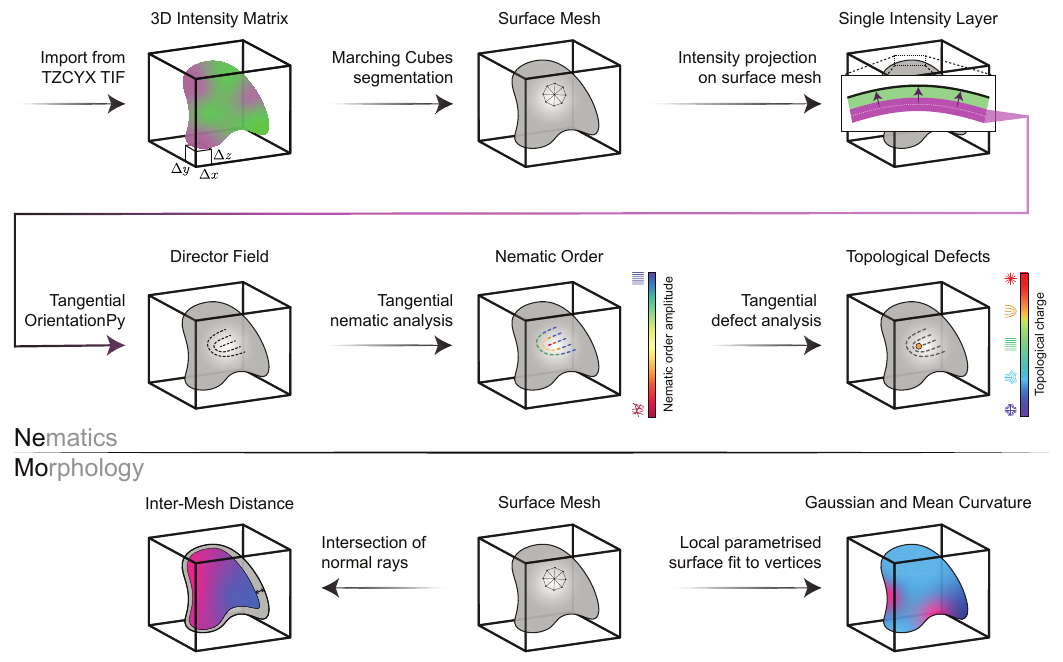}
        \caption{
            \textbf{Overview of the NEMO pipeline.}
            Starting from a TCYX TIFF file, a single 3D intensity matrix corresponding to one time point and channel is loaded into memory.
            After pre-processing the image, using a combination of Gaussian blurring and thresholding, the surfaces enclosing the binary mask are reconstructed as triangular meshes using the \textit{Marching~Cubes} method.
            Intensity signal within an interval of distances from the meshes are projected along the vertex normals on the mesh.
            From this projection, nematic director fields are extracted and averaged.
            Topological defects are identified within regions of low nematic order, and their topological charge is computed.
            Alongside nematic analysis, NEMO allows for the computation of geometrical quantities, such as surface-to-surface distance using ray casting along surface normals, and local Gaussian and mean curvatures via parametrised surface fitting.
        }
        \label{fig:overview-nemo}
    \end{figure*}

    Figure~\ref{fig:overview-nemo} summarises the NEMO pipeline workflow. The pipeline is modular, allowing users to adapt its components---from surface segmentation to defect tracking---to their specific experimental systems. Only a single z-stack from one channel and time point is loaded into memory to perform the calculations, allowing the fast processing of GB-scale files. Methods underlying the pipeline are described in more detail in Section~\ref{sec:methods}.
    \clearpage
    
    \textbf{Reconstruction of a surface as mesh.}
    
    NEMO segments the available surfaces by pre-processing the image into a binary mask using Gaussian blurring and a threshold, identifies regions above a threshold intensity  and reconstructs surfaces enclosing these regions as triangulated \textit{Trimesh} meshes \cite{trimesh} using the \textit{Marching Cubes} algorithm \cite{scikit-image}. The full mesh can then be split into its connected components, smoothed, or modified externally. The splitting functionality is important for instance to separate the apical and basal interfaces of an epithelium \cite{pernollet2026analysis}.

    \textbf{Morphological analysis of a mesh.}

    In order to characterise surface geometry, NEMO computes the local Gaussian and mean curvatures using a second-order surface fitted to a local patch of vertices, as detailed in Methods~\ref{subsec:morphological-analysis}. Basic metrics of volume and area are also provided for closed meshes in \textit{Trimesh}. Additionally, the distance separating two surfaces, such as apical and basal interfaces of an epithelium, can be computed using ray intersections along the normal vectors. Curvature and inter-layer distance can then directly be correlated with the extracted nematic fields, facilitating the study of curvature-sensing mechanisms or geometry-driven alignment.
    
    \textbf{Mesh-based projection of an intensity layer.}

    A key capability of NEMO is \textit{peeling}, the projection of intensity layers defined by their distance from the triangular mesh approximating the underlying curved surface. Each vertex of the mesh has an associated normal vector, along which the intensity is sampled and interpolated. This provides an intensity profile along the normal, allowing to identify structures away from the main surface. Intensities of ``peeled" layers are  obtained by taking the mean or maximum intensity within an interval of distances away from the main surface.

    \textbf{Extraction of tangential nematic directors.}
    
    To extract the tangential nematic director locally for a selected vertex, we define a small patch on the surface. The patch, a subset of mesh vertices, is found either by imposing a threshold distance to the central vertex or a fixed number of nearest neighbour vertices. Since the mesh resolution is typically very high in the examples considered here, only a random subset of vertices within the mesh is chosen to calculate and display the nematic field. Each selected patch is projected onto the tangential plane of the central vertex and interpolated on a regular grid to form a series of flat "micro-images". With OrientationPy, a Python analogue of Fiji's OrientationJ \cite{puspoki2016transforms, orientationpy}, the gradient structure tensor method then computes the main fibrous orientation of each micro-image as an angular field. For each patch, the central nematic angle is lifted to the corresponding tangential plane, forming a nematic director field on the curved mesh.

    \textbf{Analysis of tangential nematic order.}
    
    For a given director field, we compute the strength of local alignment and average orientation. To eliminate the need for tangential plane projections at this stage, NEMO utilises a tangential tensorial average. 
    
    First, three-dimensional tensors are constructed from the director field. They are then averaged within a patch of custom size and reprojected onto the central tangent plane. The eigenvalues and eigenvectors within the tangent plane then allows to associate to each vertex with a director, a nematic order amplitude $S$ and corresponding average orientation director $\mathbf{n}$. The alignment of nematic orientation between two layers can be obtained as further detailed in Section~\ref{subsec:tangential-nematic-analysis}.

    \textbf{Detection and analysis of topological defects.}
    
    Topological defects are identified by localising points where the local nematic order vanishes. NEMO detects the location of such point defects using an iterative scheme, as detailed in Methods~\ref{subsec:defect-detection-and-topological-charge}.
    Starting from the vertices with the lowest nematic order amplitude, other locations within a fixed geodesic distance are discarded. This is then repeated for a set number of iterations.
    While the topological charge of a defect is typically calculated via a line integral along a contour enclosing the defect, this approach assumes a flat surface. We address this limitation by reformulating the calculation for a curved surface, bypassing the need for tangential projection. As detailed in Methods~\ref{subsec:defect-detection-and-topological-charge}, the necessary correction is given by an area integral of the Gaussian curvature.
    
    \subsection{Tests on synthetic and experimental data}\label{subsec:applications-to-biological-systems}
    We validate our approach using synthetically generated images of a fibrous structure and experimental data from young \textit{Hydra}, a model organism exhibiting nematic actin organisation.

    \textbf{Synthetic image of a fibrous nematic with four comet defects on a spherical vesicle.}
    
\begin{figure*}
        \centering
        \includegraphics[width=\textwidth]{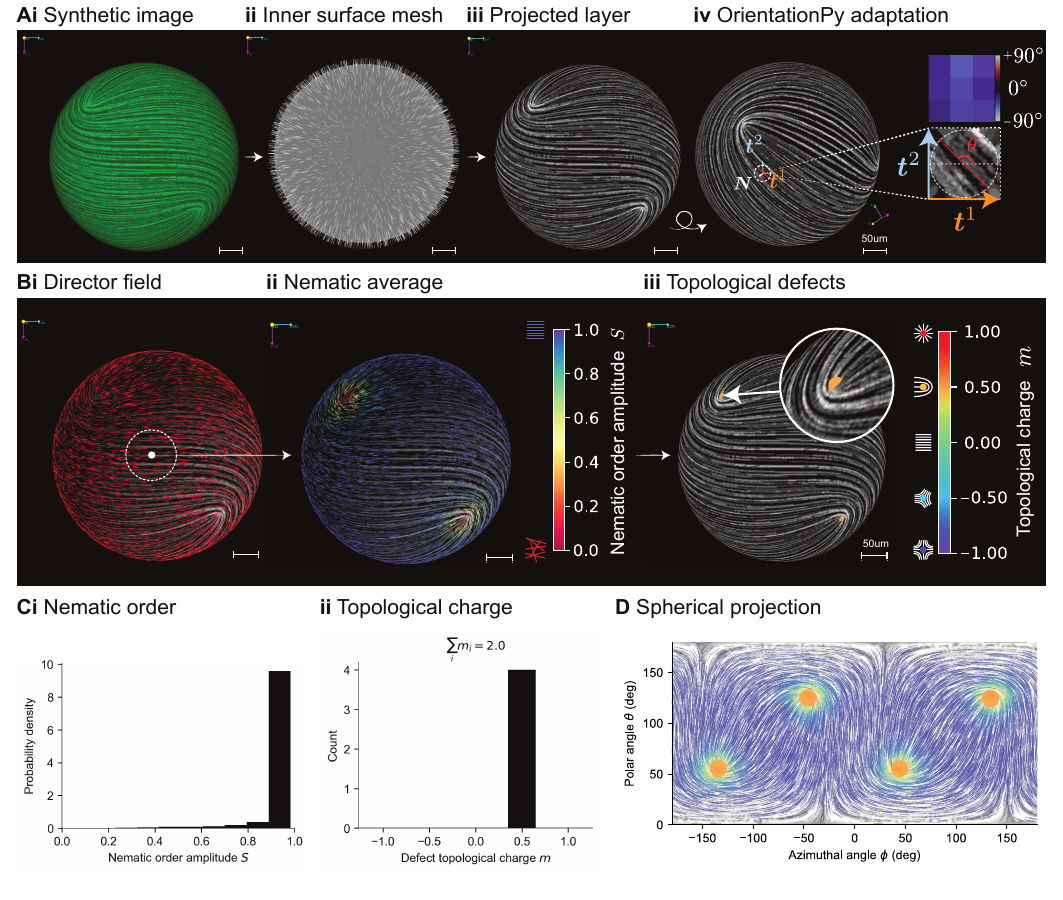}
        \caption{
            \textbf{Analysis steps taken by the NEMO pipeline to extract and analyse a tangential nematic field of a synthetic image.}
            (\textbf{A}) Synthetic image of nematic fibres on a spherical vesicle with four comet defects arranged tetrahedrally, each carrying charge $m=+1/2$ (\textbf{i}).
            Synthetically drawn fibres are contained between an inner and an outer surface, of which only the inner one is reconstructed as a triangular mesh (\textbf{ii}).
            The intensity layer containing the fibres is projected onto this surface (\textbf{iii}), and a tangential adaptation of OrientationPy computes the nematic angle in the tangential orthonormal $(\mathbf{t}^1,\mathbf{t}^2)$ basis. Insets: angular field obtained from OrientationPy (top), overlay of the central nematic angle $\theta$ (red) over the projected micro-image (bottom).
            (\textbf{B})  Nematic order amplitude $S$ and averaged director $\mathbf{n}_\text{avg}$ (\textbf{ii}) following nematic tensor averaging (\textbf{i}), and identified defects (\textbf{iii}). 
            (\textbf{C}) Nematic order (\textbf{i}) and defect topological charge  (\textbf{ii}) probability distributions.
            (\textbf{D}) 
           Representation in spherical coordinates of projected intensity signal on the vesicle surface.
            The nematic directors are coloured by the nematic order amplitude $S$ and the point defects by their topological charge, see colorbars in \textbf{B}.
        }
        \label{fig:synthetic-examples}
    \end{figure*}

    We constructed and analysed a three-dimensional image of nematically arranged fibres forming 4 $1/2$ topological defects, on a spherical vesicle with a radius set to $200 \mu m$ (Figure~\ref{fig:synthetic-examples}~\textbf{Ai}), mimicking the vesicles of Keber et al.\cite{keber2014topology}. After segmenting the surfaces, the inner mesh was isolated, reconstructing the underlying spherical vesicle (Fig.~\ref{fig:synthetic-examples}~\textbf{Aii}). 
    Fibre organisation was quantified by projecting the intensity signal arising from fibres at an interval of $[0, 5]\mu m$ away from the inner mesh, on the inner mesh (Fig.~\ref{fig:synthetic-examples}~\textbf{Aiii}).
    We then applied the tangential OrientationPy adaptation to extract a director field (Fig.~\ref{fig:synthetic-examples}~\textbf{Aiv}). This director field was then analysed using a tangential nematic tensor average (Fig.~\ref{fig:synthetic-examples}~\textbf{Bi}), yielding a measure of local alignment with the order parameter $S$ (Fig.~\ref{fig:synthetic-examples}~\textbf{Bii}). In regions of low $S$, the four topological defects were detected within $\approx 3.58 \mu m$ of their expected positions (Fig.~\ref{fig:synthetic-examples}~\textbf{Biii}).
    
    This resolution is set by the used \textit{Marching Cubes} step size (here $2 \mu m$), the spatial resolution of the extracted nematic field (only $12k$ patches were considered), and the analysis length scales ($20 \mu m$ for the extraction of the nematic directors and $50 \mu m$ for the tangential tensorial average). The defect charges were found to be $m=+1/2$, as expected (Fig.~\ref{fig:synthetic-examples}~\textbf{Cii}). Overall, we show that all four tetrahedrally positioned defects in the synthetic image can be found using a combination of mesh-based intensity projections and tangential nematic analysis (Fig.~\ref{fig:synthetic-examples}~\textbf{D}).

    \textbf{Multilayer actin organisation in \textit{Hydra}.}
    
    We used NEMO to analyse experimental data obtained from \textit{Hydra} (Figure~\ref{fig:hydra-examples}). \textit{Hydra} was chemically fixed and labelled with Alexa Fluor 555-conjugated phalloidin, a marker of actin fibres, and imaged in 3D using 2-photon microscopy. Interestingly, without any additional differential staining, we were able to independently analyse the actin alignment in the ecto- and endodermal layers, measure the distance between the endo- and ectodermal surface of the foot, and compute the curvature of the head and tentacles.

    Starting from a close-up image of a foot shown in Figure~\ref{fig:hydra-examples}~\textbf{Ai}, the ecto- and endodermal apical surface meshes were reconstructed (Fig.~\ref{fig:hydra-examples}~\textbf{Aii}) and post-processed in \textit{MeshLab} \cite{meshlab}. These meshes could then be used to project intensity layers of interest. As both ecto- and endodermal cells have an apical and basal side \cite{buzgariu2015multi}, we chose to split the tissue into four distinct layers: The ectodermal apical ($\alpha$) and basal surface ($\beta$), and the endodermal basal ($\gamma$) and apical surface ($\delta$) (Fig.~\ref{fig:hydra-examples}~\textbf{Aiii}).
    Layer $\alpha$ was obtained by projecting onto the ectodermal mesh fluorescence intensity signal coming from distances within $[0, 5]\rm \mu m$ away from the ectodermal surface, towards the center of the tissue. The layers $\beta, \gamma, \delta$ were obtained by projecting signal coming from distances away from the endodermal surface and towards the center of the tissue, within $[4.4,4.6]\rm \mu m, -[0.4, 0.6]\rm \mu m, [0, 0.2]\rm \mu m$, respectively, onto the endodermal mesh. For layers $\beta$ and $\gamma$, the range of distances for the projections was shifted for each vertex by adding 20\% of the endo-ectodermal surface distance $d$, calculated according to Eq. \ref{eq:inter-mesh-distance}.

    While the apical layers $\alpha$ and $\delta$ depict a network of cell-cell junctions with no clear nematic alignment, in $\beta$ and $\gamma$ the two basal fibrous structures seem to have a large-scale nematic alignment.
    To quantify the nematic order in both layers $\beta$ and $\gamma$ (Fig.~\ref{fig:hydra-examples}~\textbf{Bi}), we extracted the director field using a patch of size $20 \mu m$ (Fig.~\ref{fig:hydra-examples}~\textbf{Bii}) and computed the nematic average using a patch of size $50 \mu m$ (Fig.~\ref{fig:hydra-examples}~\textbf{Biii}). 

    In order to compare  the nematic field between the two basal layers, we calculated a nematic alignment coefficient between the directors of layers $\beta$ and $\gamma$ (Eq. \ref{eq:inter_layer_alignment}, Fig.~\ref{fig:hydra-examples}\textbf{C}). 
    Negative values of the alignment coefficient indicate that, as expected, the ectodermal and endodermal actin nematic order are mostly orthogonal to each other (Fig.~\ref{fig:hydra-examples}\textbf{C}).
    There are two exceptions, denoted by a hat (\textasciicircum) and an asterisk ($*$). The region denoted by the hat seems to have both layers aligned in their nematic. This is due to a projection-induced artefact, as layer $\gamma$ shown in Fig.~\ref{fig:hydra-examples}~\textbf{Bi} also contains fibres of layer $\beta$. The second region, denoted by the asterisk and corresponding to the tip of the foot, also seems to be aligned across both layers. There, the raw director field in both layers exhibits a complex pattern with decreased coherence of fibre alignment (Fig.~\ref{fig:hydra-examples}Bii). Future work could investigate the details of actin fibres ordering in that region, which overall has a topological charge $+1$.
    We also noticed the absence of actin fibres in small regions of the basal projection $\gamma$, which could be due to the quality of the actin labelling or local tissue damage (Fig.~\ref{fig:hydra-examples}Bi).
    
    Next, around the head and tentacles (Fig.~\ref{fig:hydra-examples}~\textbf{Di}), fluorescence intensity signal coming roughly from the ectodermal cells was projected on the surface mesh corresponding to the apical surface of ectodermal cells (Fig.~\ref{fig:hydra-examples}~\textbf{Dii}).
    The nematic average obtained using a patch of size $30 \mu m$ shows the circumferential alignment of cells around the tentacles (Fig.~\ref{fig:hydra-examples}~\textbf{Diii}).
    
    Finally, beyond nematic analysis, we also included examples of morphological and geometrical analysis.
    Quantifying the endoderm-to-ectoderm tissue thickness profile of the foot reveals a thicker region near the tip of the foot (Fig.~\ref{fig:hydra-examples}~\textbf{E}). Quantifying the Gaussian and mean curvatures of the head and tentacles obtained using patches of size $40 \mu m$ clearly identify regions of high Gaussian and mean curvature at the tentacle tips while the base of the tentacles exhibits negative Gaussian curvature (Fig.~\ref{fig:hydra-examples}~\textbf{Fi, Fii}).

    We hope that NEMO enables more advanced morpho-nematic studies, e.g., exploring \textit{Hydra's} dynamic actin network during normal functioning and regeneration.


    \section{Discussion}\label{sec:discussion}
    We introduce NEMO, an open-source Python-Napari framework enabling the projection of thin intensity layers, quantification of tangential nematic order, characterisation of topological defects, and morphological analysis in three-dimensional curved and multilayer biological structures. By combining mesh-based surface reconstruction with differential geometry, NEMO reduces projection-induced distortions.

    A central innovation of NEMO is the ability to \textit{peel} volumetric data by projecting intensity layers onto segmented curved surface meshes using interpolated sampling rays normal to the surface. This approach separates layers that are typically mixed in conventional projections, opening new possibilities for depth-resolved visualisation and analysis. Our application to \textit{Hydra} demonstrates this capability, revealing that actin layers—despite being highly curved and separated by only a few microns—can be computationally isolated to analyse their individual local nematic order.

    Furthermore, NEMO enables the tangential tensorial computation of nematic order and topological defect charge, regardless of the underlying surface geometry. The integration of nematic order with morphological metrics—such as local inter-surface distance and curvature—provides a more advanced framework for biophysical studies of curvature-driven or curvature-sensing alignment. Beyond one layer, computing the nematic alignment between layers, e.g. as a function of tissue depth, also allows users to study the depth dependence of a structure. Critically, the modular architecture ensures compatibility with external software at every stage of the analysis, i.e. allowing users to import custom meshes or perform more advanced analyses using custom scripts.

    Current limitations are primarily linked to the quality of the surface mesh and the sensitivity of analysis parameters. Since NEMO utilises a \textit{Marching Cubes} algorithm on blurred and masked intensity volumes, the resulting mesh is inherently sensitive to the signal-to-noise ratio and local intensity fluctuations. Inaccuracies in the mesh geometry will directly propagate into the projection quality, tangential computation of nematic order, distance between meshes, and the Gaussian and mean curvature. Such errors typically arise from inhomogeneous distances between the mesh and the actual surface, or from local high-curvature "spikes", often caused by extruding tissue cells or overexposed features on the surface of the sample. In addition, as with any orientational analysis, the resulting nematic field is heavily influenced by the chosen length scales for director extraction and tensor averaging as discussed extensively by Rembert et al.\cite{rembert2026nematic}.
    
    While NEMO offers the flexibility to tune these scales, there is an unavoidable trade-off: using a large radius for the local computation patch can compromise the validity of the local tangential approximation on highly curved surfaces. Although we provide Napari-based visualisation to validate these choices qualitatively, a formal standardised benchmarking protocol across varying noise levels remains lacking and could be addressed in further development.

    We also demonstrate that while NEMO does not need differential labelling of actin to distinguish between the endo- and ectodermal actin structures of \textit{Hydra}, the final nematic analysis result is necessarily affected by the quality of the staining.
    
    The most compelling future application of NEMO lies in coupling nematic fields to shape and flow. For shape coupling, curvature and inter-surface distance values can be directly compared to the measured nematic order on the surface. For flow coupling, one could perform particle image velocimetry (PIV) and relate velocity fields to defect locations. Beyond surface-based analysis, expanding the framework toward true 3D orientational order—addressing bulk topological defects and surface-to-bulk coupling—could allow to investigate order within dense, non-epithelial tissues.

    By making NEMO accessible to the community, we therefore aim to facilitate surface-based image analysis for biophysical studies exploring the interplay between orientational order and geometry across scales.


    \section{Methods}\label{sec:methods}
    For each analysis step shown in Fig.~\ref{fig:overview-nemo}, NEMO independently saves and loads the relevant intermediate results.
    A 3D rendering of each step is achieved using Napari, with customisable layers of, e.g., an intensity matrix, points, vectors, and triangular meshes coloured according to a scalar field.

    \subsection{Image loading}\label{subsec:image-loading}
    NEMO is compatible with multidimensional TIFF hyperstacks of dimension ($T, Z, C, Y, X$).
    Using the Python libraries \textit{TiffFile} and \textit{Zarr}, the framework virtually only loads an individual Z-stack with coordinates ($X,Y,Z$) for a specific channel ($C$) and time point ($T$). This approach allows GB-scale datasets to be processed in seconds, or to be rendered with channels and time points using Napari in minutes. The reference coordinate system for all following analyses is defined by the 3D intensity matrix using the voxel scaling ($\Delta_x,\Delta_y,\Delta_z$) and the units present in the TIFF metadata.

    \subsection{Surface reconstruction as a mesh}\label{subsec:surface-reconstruction}
    To segment the surfaces of a volumetric image as triangular meshes, NEMO implements a reconstruction algorithm based on intensity. The input image $I(x,y,z)$ is first smoothed using a 3D Gaussian filter with standard deviation $\sigma $. To account for optical anisotropy, the standard deviations along the $x$, $y$ and $z$ axes are scaled according to the ratio between the mean lateral voxel resolution $(\Delta_x, \Delta_y)$ and the axial resolution $\Delta_z$. With a Yen-based or user-defined threshold, the blurred image is transformed into a binary mask. This mask can also be supplied by the user. 
    
    From this mask, the surface is reconstructed using the \textit{Marching Cubes} algorithm \cite{scikit-image}, generating an initial triangulated mesh $\mathcal{M}$ consisting of vertices $\mathbf{V}$, faces $\mathbf{F}$ and vertex normals $\mathbf{N}$ defined at the vertices, maintained by the library \textit{Trimesh} \cite{trimesh}. To remove stair-step artifacts while preserving the total mesh volume and preventing shrinkage, we apply a Taubin smoothing step using \textit{PyVista} \cite{sullivan2019pyvista}. Since multiple surfaces may be present, such as apical and basal, or upper and lower interfaces, we offer the option to split the full mesh into two distinct sub-meshes. This is done by computing the sign of the dot product between the surface normals and their positional vector relative to the centre of mass, or compared to a custom reference axis. To remove small unwanted loose features that are segmented, the \textit{Trimesh} connected-component algorithm allows users to retain only the largest sub-mesh \cite{trimesh}. All possible meshes are saved and the user can choose which one to use in further analysis. If a perfect reference geometry is sought, the user could fit, e.g., a perfect sphere or any other kind of geometry provided by the \textit{Trimesh} library. Finally, the meshes can be exported as a Polygon File Format (PLY) file, and can be post-processed further if needed using any mesh editing software, such as \textit{MeshLab} \cite{meshlab}.

    \subsection{Projection of an intensity layer}\label{subsec:intensity-projection}
    To map a 3D intensity matrix $I(x,y,z)$ onto a PLY-format mesh, NEMO performs a vertex-wise intensity projection along the surface normals computed using \textit{Trimesh}.
    For each mesh vertex $\mathbf{v} \in \mathbf{V}$, a sampling trajectory is defined along its normal vector $\mathbf{N}$. Intensity values are sampled at $n$ discrete points using a \textit{SciPy}-based regular grid interpolator (linear or cubic). The spatial coordinates of the sampling points $\mathbf{x}_s$ are calculated as:
    \begin{equation}
        \mathbf{x}_s = \mathbf{v} + (d_{\min} + \tau)\mathbf{N}
        \label{eq:sampling_trajectory}
    \end{equation}
    where $d_{\min}$ is the initial offset from the vertex and the relative sampling depth is given by $\tau \in [0, d_{\max} - d_{\min}]$.

    This implementation allows for selective "layer peeling"; specific projection intervals relative to the mesh are set by adjusting $[d_{\min}, d_{\max}]$. The intensities collected along each trajectory are aggregated into a single scalar value per vertex using either a maximum or a mean projection, customisable by the user. Each layer is thus defined by the mesh containing the vertices, faces and normals, and a separate vertex-wise scalar field of the projected intensities.

    \subsection{Tangential nematic analysis}\label{subsec:tangential-nematic-analysis}
    To quantify orientational order on curved surfaces, we first extract the nematic directors and then compute the average local tangential nematic tensors. 

    \textbf{Definition of a local patch on the mesh.}
    
    Given a set of vertices in 2- or 3-dimensional space, neighbouring vertices for each vertex are found using a \textit{KDTree algorithm} \cite{scikit-learn}. For a given vertex, a local patch is defined using either the $k$ vertices which are nearest neighbours, or vertices that lie within a distance $r$ with respect to the central vertex. Given the large size of meshes, we do not perform the nematic analysis on every patch, but on a randomly drawn random subset of patches.

    \textbf{Local tangential projection.}
    
    To define a local orthonormal tangential basis $(\mathbf{t}^1$, $\mathbf{t}^2)$ at each vertex, we use the parallel vector transport method of Sharp et al. incorporated in the library \textit{potpourri3d} \cite{sharp2019vector}. If the mesh quality is poor, i.e. it contains loose components or holes, we instead determine $\mathbf{t}^1$ by computing the cross product of the surface normal $\mathbf{N}$ with the $\mathbf{\hat{x}}$ axis. To avoid a singularity when the normal is parallel to $\mathbf{\hat{x}}$, the $\mathbf{\hat{y}}$  axis is used.

    \textbf{Gradient structure tensor.}
    
    To determine the local orientation of anisotropic structures (e.g. fibres), we implement the standard 2D gradient structure tensor. In a tangentially projected 2D surface patch of custom radius or number of nearest neighbours, the local nematic angle is computed using OrientationPy \cite{orientationpy} and then lifted to 3D using the basis of the central vertex.

    \textbf{Definition and averaging of the nematic tensor.}
    
    For a given director field $\mathbf{n}$ on a surface with $|\mathbf{n}|=1$, the components of the uniaxial nematic order parameter tensor $q_{ij}$ in a local orthonormal basis $\mathbf{t}^1$, $\mathbf{t}^2$ read:
    \begin{equation}
        q_{ij} = n_i n_j - \frac{1}{2}\delta_{ij}
        \label{eq:q_tensor_local}
    \end{equation}
    
    As $q_{ij}$ lies in a tangential plane spanned by $\mathbf{t}^1$ and $\mathbf{t}^2$, we define the three-dimensional tensor $\tilde{\mathbf{q}}$:
    \begin{equation}
        \tilde{\mathbf{q}} = q_{ij} (\mathbf{t}^i \otimes \mathbf{t}^j)
        \label{eq:Q_lifted}
    \end{equation}
    
    To obtain an average tensor on a patch of the triangulated surface around a central vertex, we average $\tilde{\mathbf{q}}$ calculated for each vertex over the $N$ neighbouring vertices of the central vertex:
    \begin{equation}
        \langle \tilde{\mathbf{q}} \rangle = \frac{1}{N} \sum_{a=1}^{N} \tilde{\mathbf{q}}^a
        \label{eq:tan-nem-avg-lifted}
    \end{equation}
    where $\tilde{\mathbf{q}}^a$ is the tensor associated to vertex $a$.
    As a final step, we project the resulting average $\langle \tilde{\mathbf{q}} \rangle$ back onto the orthonormal tangential basis of the central vertex and we remove the trace.
    Diagonalising this tensor yields the largest eigenvalue $\lambda_{\max}$ and its corresponding eigenvector. The nematic order amplitude $S$ is then defined as $S=2\lambda_{\max}$, while the eigenvector gives the average orientation $\mathbf{n}_{\text{avg}}$.

    \textbf{Inter-layer nematic order.}
    To quantify the nematic alignment between two layers $i$ and $j$, we identify for each director $\mathbf{n}_i$ of layer $i$ the closest director $\mathbf{n}_j$ of layer $j$. Both directors are projected onto the local orthonormal tangential basis $(\mathbf{t}^1, \mathbf{t}^2)$ of layer $i$ to extract their respective orientation angles, $\theta_i$ and $\theta_j$. The inter-layer nematic alignment $c \in [-1, 1]$ is then computed as:
    \begin{equation}
        c = \cos\big(2(\theta_j - \theta_i)\big)
        \label{eq:inter_layer_alignment}
    \end{equation}
    where $c = 1$ indicates nematic alignment and $c = -1$ orthogonal alignment.

    \subsection{Defect detection and topological charge} \label{subsec:defect-detection-and-topological-charge}
    \textbf{Localisation of defects}
    
    Topological defects are identified as regions where the nematic order amplitude $S$ falls below a defined threshold. To avoid detecting multiple defects within the same low-order region, candidate defects are filtered by a minimum inter-defect geodesic distance $D_{\text{geo}}(i, j)$, defined as the shortest path length between vertices $i$ and $j$ along the surface mesh edges using the library \textit{potpourri3d} of Sharp et al. \cite{sharp2019vector}.
    
    \textbf{Topological charge on curved surfaces}
    
    We consider a patch $\mathcal{S}$ of a curved surface, with a tangential unit vector field $\mathbf{n}$ with $|\mathbf{n}|=1$, enclosed by a contour $\mathcal{C}$. The question is to calculate the topological charge $m$ within the patch.
    
    Within a flat plane with normal vector $\mathbf{e}_z$, the topological charge $m$ on a contour $\mathcal{C}$ around a defect can be calculated as an integral over the contour:
    \begin{align}
    m=\frac{1}{2\pi} \oint_{\mathcal{C}} ds ( \mathbf{n}\times (\partial_s \mathbf{n}) )\cdot\mathbf{e}_z~,
    \end{align}
    where $s$ is an arc-length coordinate over the contour, going counter-clockwise in the $(x,y)$ plane. Denoting $\mathbf{n}=(\cos\theta,\sin\theta,0)$, this can rewritten
    \begin{align}
    m=\frac{1}{2\pi}\oint_{\mathcal{C}} ds  \partial_s \theta~,
    \end{align}
    which indeed measures the total number of turns of $\theta$ over the contour.
    
    We now consider a curved surface, and follow notations of Refs. \cite{salbreux2017mechanics, salbreux2022theory}. For a defect located at point $\mathbf{X}_0$, a small contour $\mathcal{C}_{\epsilon}$ of size $\epsilon$ can be drawn around the defect. As $\epsilon$ goes to $0$, the contour lies within the tangent plane at $\mathbf{X}_0$ and the charge is obtained by
    \begin{align}
    m=\frac{1}{2\pi}\oint_{\mathcal{C}_{\epsilon}} ds ( \mathbf{n}\times (\partial_s \mathbf{n}) )\cdot\mathbf{N}~,
    \end{align}
    with $\mathbf{N}$ the normal vector to the surface at the defect point or on the contour $\mathcal{C}_{\epsilon}$, and the contour is going counterclockwise when seen from the tip of $\mathbf{N}$. We now slightly rewrite this integral
    \begin{align}
    m=\frac{1}{2\pi}\oint_{\mathcal{C}_{\epsilon}} ds^i  u_i~,
    \end{align}
    where $\mathbf{u}$ is the vector with components $u^i= ( \mathbf{n}\times (\partial^i \mathbf{n}) )\cdot\mathbf{N}$, and the vector $\mathbf{ds}$ is tangent to the contour and has length $ds$.
    
    Now considering an arbitrary contour $\mathcal{C}$ around the defect, one can show that
    \begin{align}
    \oint_{\mathcal{C}} ds^i u_i = \oint_{\mathcal{C}_{\epsilon}} ds^i u_i -\int_{\mathcal{S}_{\epsilon}} dS K~.
    \end{align}
    with $\mathcal{S}_{\epsilon}$ the patch between $\mathcal{C}$ and $\mathcal{C}_{\epsilon}$. This follows from Green's theorem \cite{riley2006mathematical} and the relation
    \begin{align}
    \epsilon^{ij} \nabla_i u_j= -K ~,
    \end{align}
    with $\epsilon_{ij}$ the Levi-Civita tensor ($\epsilon_{12}\sqrt{g}$) and $K$ the Gaussian curvature. Using $u_i=n^k\nabla_i n^l \epsilon_{kl}$, one indeed finds
    \begin{align}
    \epsilon^{ij}\nabla_i u_j=& \epsilon^{ij} \nabla_i ( n^k\nabla_j n^l \epsilon_{kl})\nonumber\\
    =& \epsilon^{ij} \epsilon_{kl} (\nabla_i  n^k)(\nabla_j n^l )  +  \epsilon^{ij} \epsilon_{kl}  n^k  \nabla_i \nabla_j n^l\nonumber\\
    =&2\det[\nabla_i n^j] + \frac{1}{2} \epsilon^{ij} \epsilon_{kl}  n^k  [\nabla_i,\nabla_j] n^l
    \end{align}
    The first term vanishes: since $\mathbf{n}^2=1$, $n_j \nabla_i n^j=0$, so that $\mathbf{n}$ is a left null vector of $\nabla_i n^j$, whose determinant then vanishes. Using the relations $\epsilon_{ij}\epsilon_{kl}=g_{ik} g_{jl} -g_{il} g_{jk}$ and $[\nabla_i,\nabla_j]n^k=C_i{}^k C_{l j}n^l - C_{il} C_j{}^k n^l$ \cite{salbreux2022theory}, one then obtains:
    \begin{align}
    \epsilon^{ij}\nabla_i u_j=C_i{}^j C_{mj} n^i n^m -C_i{}^i C_{mj} n^m n^j~.
    \end{align}
    Using a local orthonormal basis with one tangent vector aligned with $\mathbf{n}$, one can verify that this scalar is equal to $-K=-\det(C_i{}^j)$.
    
    When $\epsilon \rightarrow 0$, we then obtain the expression for the charge evaluated on an arbitrary contour around the defect:
    
    \begin{align}
    \label{eq:final_defect_charge}
    m = \frac{1}{2\pi}\oint_{\mathcal{C}_{\epsilon}} ds^i u_i =\frac{1}{2\pi} \oint_{\mathcal{C}} ds^i u_i  +\frac{1}{2\pi}\int_{\mathcal{S}} dS K ~.
    \end{align}

     Evaluated over $N$ directors of the contour, the first integral of the right hand-side of Eq.~\ref{eq:final_defect_charge} is discretised as:
    \begin{equation}
        \frac{1}{2\pi} \oint_{\mathcal{C}} ds^i u_i \approx \frac{1}{2\pi}\sum_{i=1}^{N}[\mathbf{n}_{i}\times (\mathbf{n}_{i+1}-\mathbf{n}_{i})]\cdot\mathbf{N}_{i} ~,
        \label{eq:winding-discrete}
    \end{equation}
    where head-to-tail symmetry is accounted for by aligning neighbouring directors using an iterative flipping $\mathbf{n}_{i+1}$ based on the sign of the dot product with $\mathbf{n}_{i}$.
    
    The second integral is implemented by summing over the Gaussian curvature of the patch, which is obtained by fitting a second-order surface as described further in \ref{subsec:morphological-analysis}.

    
    


    \subsection{Morphological analysis}\label{subsec:morphological-analysis}
    Since meshes can be noisy and can contain more than a million vertices, the following morphological quantities are computed only on a random subset of vertices. To interpolate the computed subset to the full mesh, we use an inverse-distance weighted average from the $k$ nearest computed values:
    \begin{equation}
        \hat{v}_i = \frac{ \displaystyle\sum_{j=1}^k d_{ij}^{-1} v_{n_{ij}} }{ \displaystyle\sum_{j=1}^k d_{ij}^{-1} }\label{eq:mesh-interpolation}
    \end{equation}
    where $\hat{v}_i$ is the interpolated value at vertex $i$, $v_{n_{ij}}$ is the known value at the $j$-th nearest neighbour of vertex $i$, $d_{ij}$ is the distance between vertex $i$ and neighbour $j$, and $k$ is the number of nearest neighbours.
    \\
    
    \textbf{Gaussian and mean curvature.}
    
    At each sampled vertex $\mathbf{V}_{0}$, we estimate the local curvature by fitting a second-order surface to its neighbouring vertices. Let $\mathbf{N}_{0}$ be the vertex normal, and define orthonormal tangential basis vectors $\mathbf{t}_{1}$, $\mathbf{t}_{2}$ such that $\{\mathbf{t}_{1},\mathbf{t}_{2},\mathbf{N}_{0}\}$ forms a local frame.
    
    For each neighbour $\mathbf{V}_{j}$, we compute:
    \begin{equation*}
        \xi_{j}=(\mathbf{V}_{j}-\mathbf{V}_{0})\cdot\mathbf{t}_{1}, \, \eta_{j}=(\mathbf{V}_{j}-\mathbf{V}_{0})\cdot\mathbf{t}_{2}, \, z_{j}=(\mathbf{V}_{j}-\mathbf{V}_{0})\cdot\mathbf{N}_{0}
    \end{equation*}
    
    The local height is approximated by:
    \begin{equation}
        z_{j} \approx a~\xi_{j}^{2} + b~\eta_{j}^{2} + c~\xi_{j}\eta_{j}\label{eq:curv-surf-fit}
    \end{equation}
    where $a$, $b$, and $c$ are obtained from a least-squares fit.
    The curvature tensor of the surface is then given by: 
    \begin{equation}
        \mathbf{C} = \begin{pmatrix}
                         -2a & -c \\ -c & -2b
        \end{pmatrix}\label{eq:curvature-tensor}
    \end{equation}

    Using the local metric tensor $g_{ij}=\mathbf{t}_{i}\cdot\mathbf{t}_{j}$ and the mixed curvature tensor $C_{i}^{j}=C_{ik}g^{kj}$, we compute:
    \begin{equation}
        H = \frac{C_{k}^{k}}{2}, \quad K = \frac{\det C_{ij}}{\det g_{ij}}\label{eq:curv-mean-gaussian}
    \end{equation}
    where $H$ and $K$ are the mean and Gaussian curvatures at vertex $\mathbf{V}_{0}$, respectively.
    
    \textbf{Distance between surfaces.}
    
    Given a set of vertices and their normals, we trace rays along the normals from surface 1 to surface 2 using the library \textit{Trimesh} \cite{trimesh}. For a vertex at $\mathbf{V}_i$ with normal $\mathbf{N}_i$, we determine the intersection $\mathbf{V}_i^*$ on mesh 2 along the ray $\mathbf{x}_i + t\mathbf{N}_i$:
    \begin{equation}
        d_i = (\mathbf{V}_i^\ast - \mathbf{V}_i) \cdot \mathbf{N}_i \label{eq:inter-mesh-distance}
    \end{equation}
    where $d_i$ is the distance projected along the normal direction.

    \subsection{Synthetic image generation}\label{subsec:synthetic-image-generation}
    To validate the capability of the NEMO framework to resolve tangential nematic order and extract topological defects on curved surfaces, we constructed a synthetic volumetric image of an analytical filamentous nematic film on a spherical vesicle of radius $R$ with four equidistant $m = +1/2$ comet defects located at the vertices of a regular tetrahedron on the sphere. This example mimics the vesicles of Keber et al. \cite{keber2014topology}. 

    To establish a continuous coordinate framework for the director field across the entire sphere, we map the surface onto the complex plane $\mathbb{C}$ via stereographic projection from the South Pole $(0,0,-1)$. A spatial vertex coordinate $\mathbf{x} = (x, y, z)^T \in S^2$ is mapped to its corresponding planar complex coordinate $w = u + \mathrm{i}v$ via:
    \begin{equation}
        w = \frac{x + \mathrm{i}y}{1 + z}
        \label{eq:stereographic_projection}
    \end{equation}
    
    Let $w_k \in \mathbb{C}$ denote the planar positions of the four defect cores mapped under Eq.~\ref{eq:stereographic_projection}. In this flattened space, the local nematic orientation angle $\psi(w)$ is determined by the linear superposition of the defects' phase contributions:
    \begin{equation}
        \psi(w) = \frac{1}{2} \sum_{k=1}^{4} \text{arctan2}\big(\text{Im}(w - w_k), \text{Re}(w - w_k)\big)
        \label{eq:synthetic_psi}
    \end{equation}
    We lift this planar angle field back onto the tangent planes of the sphere to obtain a director field $\mathbf{n}(\mathbf{x})$.
    
    Discrete filaments are generated by tracing bidirectional streamlines along $\mathbf{n}(\mathbf{x})$. To convert these geometric trajectories into realistic fluorescence microscopy data, curves are rasterized onto a three-dimensional voxel grid of dimensions $G \times G \times G$ with a voxel size of $1\,\mu\mathrm{m}$. Points along each trace on the unit sphere are mapped to voxel coordinates $\mathbf{x}^*$ by scaling and translation:
    \begin{equation}
        \mathbf{x}^* = R\mathbf{x} + \mathbf{c}
    \end{equation}
    where $\mathbf{c} = (G/2, G/2, G/2)^T$ is the centre of the volume and $R$ the radius of the sphere.

    Voxels within a set distance from the surface are assigned a constant background intensity, and voxels intersected by the filament curves are assigned a foreground intensity. After combining these features into the intensity grid, the volume is blurred by convolving it with an isotropic 3D Gaussian kernel. The values are then multiplied by 255 and clipped to the interval $[0, 255]$ to replicate a realistic experimental 3D image.
    
    \subsection{\textit{Hydra} culture and imaging}\label{subsec:hydra-culture-and-imaging}
    \textbf{Hydra culture.} 
    
    Wild-type \textit{Hydra} from the AEP strain were cultured in standard \textit{Hydra} culture medium (HM; 1~mM NaHCO\textsubscript{3}, 1~mM CaCl\textsubscript{2}, 0.1~mM MgCl\textsubscript{2}, 0.1~mM KCl, and 1~mM Tris-HCl, pH~7.7) at 18~°C. Animals were fed three times per week with live \textit{Artemia} nauplii. To select small individuals for imaging, young animals were isolated from the culture during the peak budding window, specifically $\sim$30~h following the last feeding cycle.
    
    \textbf{Hydra fixation and staining.}
    
    \textit{Hydra} were relaxed in 2\% urethane in \textit{Hydra} medium (HM) for 1~min and fixed in 4\% paraformaldehyde (PFA) in HM for 1~h at room temperature. Samples were permeabilised with 0.1\% Triton X-100 in PBS (PBT) and washed three times for 5~min each. For staining, samples were incubated with Alexa Fluor 555-conjugated phalloidin (200~U/ml; Invitrogen) diluted 1:400 in 0.1\% Triton X-100 in PBS, and incubated overnight at 4~°C with gentle shaking. Subsequently, samples were washed three times in PBS (5~min each). Stained specimens were immediately mounted in 2\% low gelling agarose for fluorescence imaging.

    \textbf{Two-photon imaging}.
    
    F-actin structures were resolved using an upright Leica SP8 DIVE FALCON multiphoton microscope configured with high-sensitivity hybrid (HyD) detectors. Two-photon excitation of the fluorophore was driven by a tunable ultrafast laser operating at 960~nm. Optical sections were acquired with a 25X HC FLUOTAR L water-immersion objective (NA 0.95). Signal decay across $z$ was corrected by dynamically modulating the excitation power and $z$-stacks were captured at 1~$\mu$m intervals.


    \section{Data availability}\label{sec:data-availability}
    The raw images, processed data, selected analysis parameters, and raw analysis results generated during this study are available from the corresponding author upon reasonable request.


    \section{Code availability}\label{sec:code-availability}
    All NEMO modules and notebooks are made publicly available on our repository: 
    
    \href{https://github.com/unige-biochem/nemo}{https://github.com/unige-biochem/nemo}


    \section{Use of Artificial Intelligence}\label{sec:use-of-ai}
     Claude by Anthropic and Gemini by Google were used during the preparation of this work.
     This includes assistance with the theoretical work on topological charge on curved surfaces, the optimisation of NEMO scripts, the development of code for the generation of the synthetic image, and manuscript editing. 

\clearpage
    \bibliographystyle{unsrt}
    \bibliography{main}

    \begin{figure*}
        \centering
        \includegraphics[width=\textwidth]{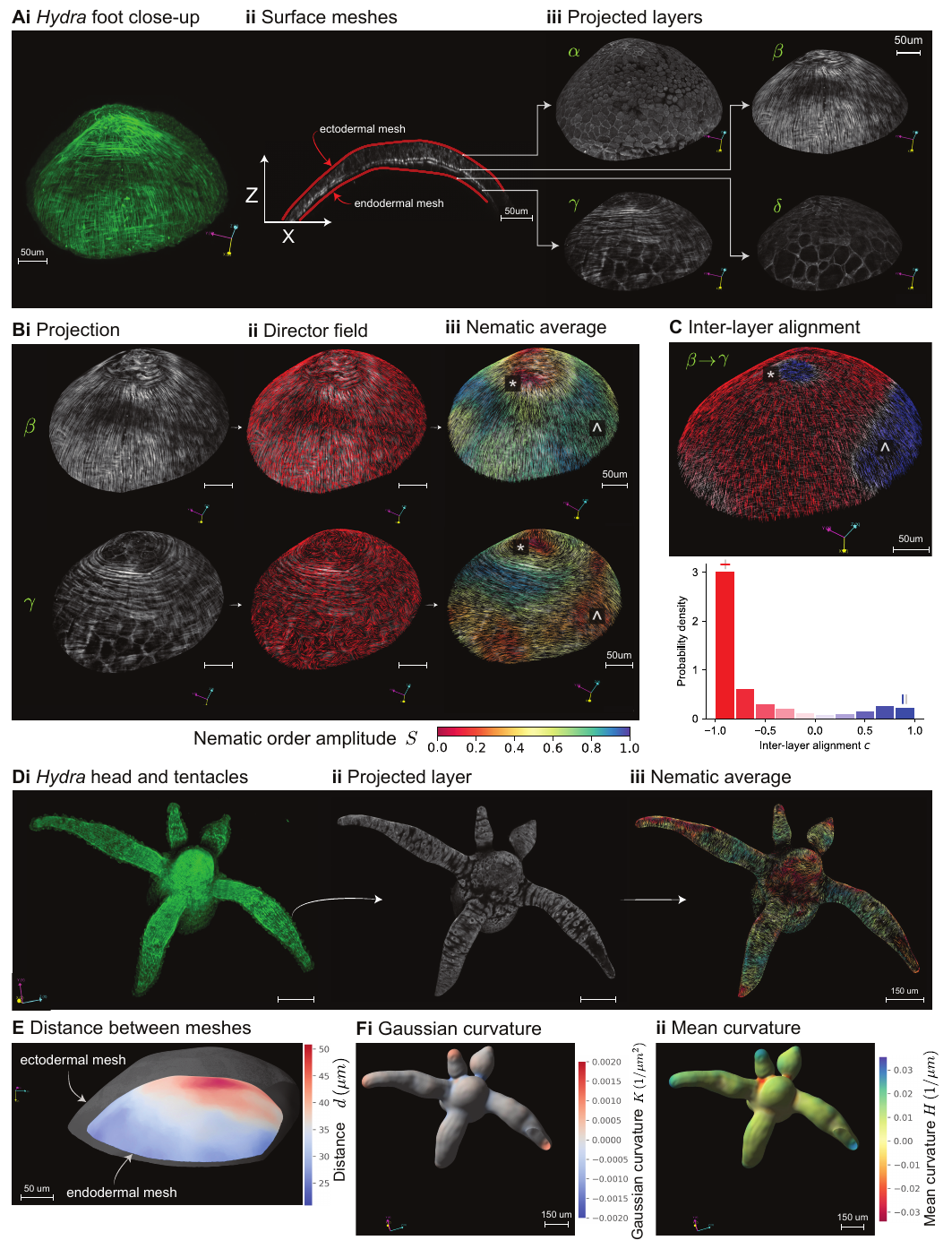}
        \caption{\textbf{Application of NEMO to the endo- versus ectodermal nematic order in \textit{Hydra} using actin labelling.} \textit{(See caption on next page)}}
        \label{fig:hydra-examples}
    \end{figure*}

    \begin{figure*}
        \ContinuedFloat
        \centering
        \caption*{
            \textbf{Application of NEMO to the endo- versus ectodermal nematic order in \textit{Hydra} using actin labelling.}

            (\textbf{A}) 3D 2-photon imaging of a \textit{Hydra} foot (\textbf{i}) with overlaid ecto- and endodermal meshes on the middle XZ slice (\textbf{ii}). Four intensity layers are projected (\textbf{iii}), revealing the apical network of cell-cell junctions (layers $\alpha$ and $\delta$) and basal fibres (layers $\beta$ and $\gamma$) . 
            (\textbf{B}) Nematic analysis of basal projections $\beta$ and $\gamma$ (\textbf{i}), the extracted director field (\textbf{ii}, $20\,\mu m$ patch size) and nematic average (\textbf{iii}, $50\,\mu m$ patch size). 
            (\textbf{C}) Inter-layer alignment comparison between layers $\beta$ and $\gamma$ demonstrating predominantly orthogonal nematic alignment. Exceptions include a projection-induced artifact (\textasciicircum)
            and the tip of the foot region with reduced nematic order ($*$).
            (\textbf{D}) Head and tentacles (\textbf{i}) projected onto the highly curved ectodermal surface (\textbf{ii}), where the nematic average (\textbf{iii}, $30\,\mu m$ patch) highlights circumferential alignment. 
            (\textbf{E}-\textbf{F}) Morphological and geometric metrics showing 3D endo-to-ecto tissue thickness in the foot (\textbf{E}), alongside Gaussian (\textbf{Fi}) and mean (\textbf{Fii}) curvatures of the head and tentacles ($40\,\mu m$ patch size).
        }
    \end{figure*}
\end{document}